\documentclass{ws-procs975x65}

\usepackage{latexsym}
\usepackage{amssymb}
\usepackage{amsmath}
\usepackage{xspace}
\usepackage{subfigure}
\usepackage{psfrag}
\usepackage{graphicx}

\usepackage[dvips]{epsfig}
 \DeclareMathOperator{\tg}{tg}
 
\DeclareMathOperator{\Sp}{Sp}

\newcommand{\ds}{\displaystyle}

\newcommand{\cp}{{\cal{P}}}

\newcommand{\vk}{\varkappa}

\newcommand{\be}{\begin{equation}}
\newcommand{\ee}{\end{equation}}
\newcommand{\ba}{\begin{align}}
\newcommand{\ea}{\end{align}}
\newcommand{\bea}{\begin{eqnarray}}
\newcommand{\eea}{\end{eqnarray}}
\newcommand{\bw}{\begin{widetext}}
\newcommand{\ew}{\end{widetext}}

\newcommand{\e}{{\rm e}}

\newcommand{\nn}{\nonumber}
\newcommand{\zt}{\dot{z}}
\newcommand{\ztt}{\ddot{z}}

\newcommand{\cd}{,\! \,}
\newcommand{\un}{\, ^1 \!}
\newcommand{\nul}{\, ^0\!}

\newcommand{\pp}{\, ... \,}

\newcommand{\eff}{\mathrm{eff}}

\newcommand{\Hh}{h}

\newcommand{\dq}{d^{{\kern 1pt}4} {\kern -1pt}q}

\newcommand{\kn}{{\kern 1pt}}
\newcommand{\akn}{{\kern -1pt}}

\newcommand{\vph}{\vphantom{{c'}^M}}
\newcommand{\vp}{\vphantom{{c}^M}}

\begin{document}

\title{Gravitational Bremsstrahlung from Massless-particle Collisions}
%%%%%  Authors  %%%%
\author{Pavel
Spirin$^{1,2}$ and Theodore\,N.\,Tomaras$^{1}$
%\footnote{E-mails: pspirin@physics.uoc.gr, tomaras@physics.uoc.gr}
}
%\thanks{\tt salotop@list.ru}}
\address{\parbox{11cm}{\noindent\rule{0cm}{0.4cm}{}$^{1}$\,Institute of Theoretical and
Computational Physics, Department of Physics,  \\ \phantom{$^{1}$}\,University of
Crete, 70013 Heraklion, Greece;
\\ {}$^{2}${}Department
of Theoretical Physics, Faculty of Physics, \\ \phantom{$^{2}$}\,M.\,V.\,Lomonosov Moscow State
University,  119911 Moscow, Russia.}}

%\pacs{11.27.+d, 98.80.Cq, 98.80.-k, 95.30.Sf}
\date{\today}

\begin{abstract}
The angular and frequency characteristics of the gravitational
radiation emitted in collisions of massless particles is studied
perturbatively in the context of classical General Relativity for
small values of the ratio $\alpha\equiv 2 r_S/b$ of the
Schwarzschild radius over the impact parameter. The particles are
described with their trajectories, while the contribution of the
leading nonlinear terms of the gravitational action is also taken
into account. The old quantum results are reproduced in the zero
frequency limit \mbox{$\omega\ll 1/b$}. The radiation efficiency
$\epsilon \equiv E_{\rm rad}/2E$ {\it outside} a narrow cone of
angle $\alpha$ in the forward and backward directions with respect
to the initial particle trajectories is given by \mbox{$\epsilon
\sim \alpha^2$} and is dominated by radiation with characteristic
frequency \mbox{$\omega \sim {\mathcal O}\kn(1/r_S)$}. The
comparison with previous works and the known literature is
presented.
\end{abstract}

%\tableofcontents
\section{Introduction}
\label{intr}

The problem of gravitational radiation in particle collisions has
a long history and has been studied in a variety of approaches and
approximations. The problem of gravitational radiation in {\it
massless-}particle collisions is worth studying in its own right
and has attracted the interest of many authors in the past as well
as very recently. Apart from its obvious relevance in the context
of TeV-scale gravity models with large extra dimensions
\cite{GKST}{}, it is very important in relation to the structure of
string theory and the issue of black-hole formation in
ultra-planckian collisions \cite{Dvali}{}. Nevertheless, to the best
of our knowledge, complete understanding of all facets of the
problem is still lacking. The emission of radiation in the form of
{\it soft} gravitons was computed in\,\cite{Weinberg} in the
context of quantum field theory, but in that computation the
contribution of the non-linear graviton self-couplings i.e. the
stress part of the energy-momentum tensor, was argued to be
negligible. The result of the quantum computation for
low-frequency graviton emission was reproduced by a purely
classical computation in\,\cite{Wbook}{}. More recently, a new
approach was put forward for the computation of the
characteristics of the emitted radiation \cite{GV}, based on the
Fraunhofer approximation of radiation theory.

 The purpose here is to extend the method used in\,\cite{GST} to the
study of gravitational radiation in collisions of massless
particles with center-of-mass energy $2E$ and impact parameter
$b$. The formal limit $m\to 0$ (or equivalently $\gamma_{\rm cm}
\to \infty$ for the Lorentz factor) of the massive case leads to
nonsensical answers for the radiation efficiency, i.e. the ratio
$\epsilon\equiv E_{\rm rad}/2E \sim (r_S/b)^3 \gamma_{\rm cm}$ of
the radiated to the available energy, the characteristic radiation
frequency $\omega\sim \gamma^2_{\rm cm} /b$, or the characteristic
emission angle $\vartheta\sim 1/\gamma_{\rm cm}$. We study
classically the gravitational radiation in the collision of
massless particles using the same perturbative approach as in
\cite{GST}{}.

 This presentation is mainly based on results obtained in\,\cite{ST}{}.

\section{Scattering}\label{eom}

The action describing the two massless particles and their gravitational interaction reads
\begin{equation} \label{action}
S=- \frac{1}{2} \sum \int  e (\sigma )\, g_{
\mu\nu}\!\left(z(\sigma)\vp\right)\dot{z}^{\mu}(\sigma)\,
\dot{z}^{\nu}(\sigma) \,d\sigma  -\frac{1}{\vk^2}\int R
\sqrt{-g}\, d^{\kn 4}\akn x\,,
 \end{equation}
where $e(\sigma)$ is the einbein of the trajectory $z^\mu(\sigma)$
in terms of the corresponding affine parameter $\sigma$,
$\vk^2=16\pi G$ and the summation is over the two particles. We
will be using unprimed and primed symbols to denote quantities
related to the two particles. For identical colliding particles in
the center-of-mass frame $ e=\sqrt{s}/2=E\,,$ with $E$ the energy
of each colliding particle.

Thus, the particles move on null geodesics, while variation of $z^{\mu}$ leads to the particle equation of motion:
\begin{align}
\label{peom}
 \frac{d}{d\sigma } \left(g_{\mu\nu}\zt^\nu\right)
 = \frac{1}{2}\,g_{\lambda\nu,\kn\mu} \zt^\lambda  \zt^\nu
\end{align}
and similarly for $z'^\mu$. The particle energy-momentum is
defined by $T^{\mu\nu} \equiv (-2/\sqrt{-g}) \,\delta S/\delta
g_{\mu\nu}$\,. At zeroth order in the gravitational interaction,
the space-time is flat and the particles move on straight lines
with constant velocities, i.e.\kn\footnote{The upper left index on
a symbol labels its order in our perturbation scheme.} $ \nul
g_{\mu\nu}=\eta_{\mu\nu} \,; \qquad \nul \dot{z}^\mu \equiv
u^\mu=(1,0, 0,1)\,, \qquad\nul \dot{z}'^\mu \equiv
u'^\mu=(1,0,0,-1)\,.$  Given that $\nul \,T_{\mu\nu}$ is
traceless, the perturbation $h_{\mu\nu}$ satisfies for each
particle separately the equation $
\partial^2 h_{ \mu\nu}=-\vk \nul\, T_{ \mu\nu}\,,
$
whose solution in coordinate representation is the Aichelburg --
Sexl metric
 \begin{align}
& h_{\mu\nu}(x) =   -\vk {\kern 1pt}e{\kern 1pt} u_{\mu} u_{\nu}
{\kern 1pt} \delta(t-z)\, \Phi(|{\bf r}-{\bf b}/2|) \, \nn
\end{align}
where $\Phi$ is the $2-$dimensional Fourier transform of $1/q^2$:
$ \Phi(r)   = - ({1}/{2\pi}) \ln ( {r}/{r_0})$ with $r_0$   an arbitrary constant length parameter.

Write for the metric $g_{\mu\nu}=\eta_{\mu\nu}+\vk
(h_{\mu\nu}+h'_{\mu\nu})$ and substitute in (\ref{peom}) to obtain
for the first correction of the trajectory of the unprimed
particle the equation
\begin{align}
\label{peom1} \un \ztt_{\mu} ( \sigma )
=-\vk\kn\Bigl(h'_{\mu\nu,\lambda}-\frac{1}{2}\,h'_{\lambda\nu,\kn\mu}\Bigr)\nul\zt^\lambda\kn
\nul\zt^\nu.
\end{align}
The interaction with the self-field of the particle has been
omitted and $h'_{\mu\nu}$ due to the primed particle is evaluated
at the location of the unprimed particle on its unperturbed
trajectory.

Integrating (\ref{peom1}) one obtains   the trajectory  in
components:
\begin{align}
\label{ddd1} & \un z^{0}( \sigma )= \frac{1}{2}\, e \vk^2
\Phi(b)\, \theta( \sigma ) =-\un z^{z}( \sigma )\,, \qquad\qquad
\un z^{x}( \sigma )=e \vk^2 \Phi'(b)\, \sigma {\kern 1pt}\theta(
\sigma )\,.
\end{align}
which vanish for all \mbox{$\sigma <0$}. Indeed, the massless
particle trajectories should remain undisturbed before the
collision. Thus, we reproduce the leading order expressions of the
two well-known facts \cite{Dray87} about the geodesics in an
Aichelburg-Sexl metric, namely: (i) the {\it time delay} at the
moment of shock equal to $\Delta t =   e \vk^2 \Phi(b)=8{\kern 1pt}
GE\,\ln ({b}/{r_0})\,;$ (ii)  the {\it refraction} caused by the
gravitational interaction by an
    angle
$ \alpha =   e \vk^2  \, |\Phi'(b)|= {8{\kern 1pt}GE}/{b} $ in the
direction of the center of gravity.

\section{Radiation amplitude}

The gravitational wave source has two parts. One is the particle
energy-momentum contribution, localized on the accelerated
particle trajectories.

  The direct particle
contribution to the source of radiation is called  ``local",
because it is localized on the particle trajectories. The first
order term in the expansion of $T_{\mu\nu}$  is
 \begin{align}
\label{Tmn1}
\un T_{\mu\nu}(x)= e\int  d\sigma \, & \left[ 2\un \zt_{(\mu}
u_{\nu)} +2 \vk u^{\lambda}h'_{\lambda
(\mu} u{\vphantom{h'_\lambda}}_{\nu)} - u_\mu u_\nu (\!\un {z}\cdot
\partial) \right] \delta^4 \!\!{\kern 1pt}\left(x-\! \! \nul z(\sigma )\right)\, ,
\end{align}
where $z^\mu$ is evaluated at $\sigma $ and $ h'_{\mu\nu}$ is
evaluated at $\nul z^\mu(\sigma ) $.

In Fourier space one obtains effectively
\begin{align}
\label{T1eff}
 \un T_{\mu\nu}=-2 {\kern 1pt} e^2 \vk^2 \e^{i(kb)/2}
 \biggl[ -  \Phi(b) \,u_{\mu} u_{\nu} +
  \frac{ (ku')\,\Phi(b)}{2 {\kern 1pt} (ku)}\,u_{\mu}u_{\nu} +i\frac{\Phi'(b)  \, \sigma^{(u)}_{\mu\nu}}{b\,  (k
u)^2 } \biggr]\,.
\end{align}
  The contribution to the source
at second-order coming from the expansion of the Einstein tensor,
reads \cite{GST}
\begin{align}
&S_{\mu\nu} =\, {\Hh}_\mu^{\lambda \cd \rho }\left(\vp\Hh_{\nu \rho
\cd \lambda } - \Hh_{\nu \lambda \cd \rho }\right) +\Hh^{\lambda \rho
}\left(\vp\Hh_{\mu \lambda \cd \nu \rho }+ \Hh_{\nu \lambda \cd \kn\mu \rho
}- \Hh_{\lambda \rho \cd \mu\nu}- \Hh_{\mu\nu
\cd \lambda \rho }\right) -  \nn\\
  & \quad -\frac{1}{2}\, \Hh^{\lambda \rho }{}_{\!\!\cd\kn \mu } \Hh_{\lambda \rho  \cd
\nu }\akn -\akn \frac{1}{2}\,\Hh_{\mu\nu} \partial^{\kn 2} \Hh \akn+\akn
\frac{1}{2}\,\eta_{\mu\nu}\Bigl(2\Hh^{\lambda \rho }\partial^{\kn 2}
\Hh_{\lambda \rho }\akn-\akn\Hh_{\lambda \rho  \cd \sigma} \Hh^{\lambda
\sigma \cd \rho }\akn+\akn\frac{3}{2}\, \Hh_{\lambda \rho  \cd \sigma}
\Hh^{\lambda \rho \cd \sigma}\akn\Bigr).\nn
\end{align}

Upon substitution of $h_{\mu\nu}$ and $\un z^\mu(\sigma)$ of the
previous section in the above expression we obtain for the Fourier
transform of $S_{\mu\nu}$
\begin{align}
S_{\mu\nu}(k) = \vk^2 e^2 \e^{i (kb)/2} & \left[ \vph (ku')^2 u_\mu
u_\nu  J+ (ku)^2 u'_\mu  u'_\nu  J+4{\kern 1pt} J_{\mu\nu}
+4{\kern 1pt} (ku')\,u_{(\mu } J_{\nu )}- \right. \nn
\\ &\left.-4{\kern 1pt}
(ku)\,u'_{(\mu } J_{\nu )} ++2 {\kern 1pt} u_{(\mu } u'_{\nu
)}\left(2{\kern 1pt} (kJ) -(ku)(ku')\,J-2\Sp J \vph\right) \right]
\nonumber
\end{align}
in terms of the integrals ($l=0, 1, 2$)
\begin{align}
\nonumber J_{\mu _1\,\pp \mu _l}(k)\equiv \frac{1}{(2\pi)^{2}}
\int \frac{ \delta(qu')\, \delta(ku-qu)\, \e^{- i(qb)}}{q^2
(k-q)^2} \;q_{\mu _1}\pp q_{\mu _l} \; \dq\,.
\end{align}

\bigskip

\textbf{Total radiation amplitudes.}  Defining in the
center-of-mass frame the radiation wave-vector by $k^{{\kern
1pt}\mu} = \omega{\kern 1pt} (1, {\bf n}) = \omega{\kern 1pt}
(1,\sin \vartheta \cos\varphi,\sin \vartheta \sin\varphi, \cos
\vartheta ) $ and contracting  with the two polarizations (defined
as usual), we obtain the final (finite for \mbox{$\vartheta>0$}) expressions for the source
of the gravitational radiation separately for the two
polarizations (here $\hat{K}_{\nu}( \zeta ) \equiv \zeta^{\nu} {K}_{\nu}( \zeta ) $, $\zeta=\omega b \sin\vartheta \sqrt{x(1-x)}$):
\begin{align}
\label{tau+} &\tau_{+}(k)  =\frac{16 {\kern 1pt}G  e^2}{\sqrt{2}
}\, \int_0^1 dx \,\e^{-i (k b) x}   \left[ - {K}_{0}( \zeta
)+\sin^2\!{\varphi}\, \hat{K}_{1}( \zeta ) \right],\\
&
 \tau_{\times}(k)  =-\frac{16{\kern 1pt} G e^2}{\sqrt{2}}\, \sin \varphi\,
\!\!\int_0^1 dx \,\e^{-i (k b) x}  \biggl[ 2i \frac{\hat{K}_{2}(
\zeta )-\hat{K}_{1}( \zeta )}{ \omega b \sin \vartheta}  +  (2x-1)
\cos   \varphi  \,\hat{K}_{1}( \zeta ) \biggr]\,. \nn
\end{align}

\section{Characteristics of the emitted radiation}
The emitted radiation frequency spectrum and of the total emitted
energy are obtained from (sum over the two polarizations):
 \begin{align}
 \label{fr_di}
\frac{d E_{\rm rad}}{d\omega \,d\Omega}=\frac{ G }{2\pi^{2}}
\,\omega^{2} \sum_{\cp} | \tau_{\cp}|^2\,.
\end{align}

\begin{figure}
 \begin{center}
 %\psfrag{a}{\small $\ds \frac{dE_{\rm rad}}{d\omega}\sim$}
\includegraphics[width=11cm]{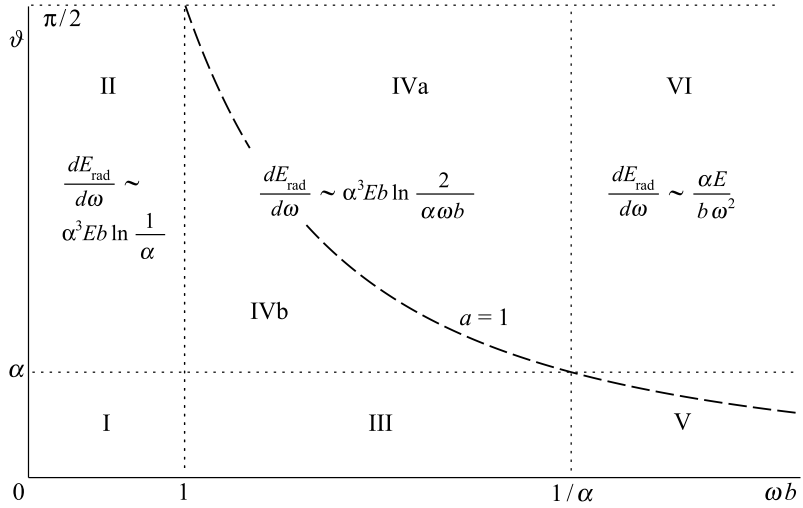}  \caption{The characteristic angular and
frequency regimes ($a\equiv \omega b \sin \vartheta$).}
\label{table}
 \end{center}
\end{figure}

It will be convenient in the sequel to treat separately the six
angular and frequency regimes shown in Fig.\,\ref{table}.

\medskip

\textbf{Zero-frequency limit.} In the low-frequency regime
($\omega \to 0$) the amplitude $\tau_{\times} $ dominates and has
the form
\begin{align}
\label{scalampl2dd}
 \tau_{\times} \simeq - \frac{16 \sqrt{2} i{\kern 1pt} G E^2  \sin \varphi}{\omega b \sin \vartheta}\,
 ,
\end{align}
while $\tau_{+}$ is finite (as $\vartheta \to 0$) and gives subleading contribution to
\eqref{fr_di}. The angular integration leads to $d E_{\rm rad
}/d\omega = (2^8 G^3 E^4/ \pi b^{\kn 2}) \int
d\vartheta/\sin\vartheta$, which diverges and implies that our
formulae are not valid for $\vartheta$ close to zero.

We cannot trust our formulae in regime I and should repair
them. A quick way to do it, is to impose a small-angle cut-off
$\vartheta_{\min}=\alpha$ on the
$\vartheta-$integration, so as to obtain for $\ds \left.dE_{\rm
rad}/d\omega \vp\right|_{\omega=0}$ the value computed quantum mechanically
in\,\cite{Weinberg,Wbook}{}:
\begin{align}
\label{DEMD_ef1}
\left( \frac{dE^W}{d\omega} \right)_{\!\omega=0}& = \frac{4}{
 \pi} \,  G  |t| \ln \frac{s}{|t|}=  \frac{G   \alpha^2 s}{
 \pi}\,   \ln \frac{4}{ \alpha^2} = \frac{E\kn b}{\pi}\, \alpha^3 \ln\frac{2}{\alpha} \,, \qquad s=4E^2\,,
\end{align}
which  agrees with our expression for $ \vartheta> \alpha \,. $
Thus,  in regime II we use our formula, which is identical to
Weinberg's. Now for any frequency we have to compare our matter part, which makes the $\vartheta-$integration divergent, with the Weinberg's:
\begin{align}
\label{GWM_Fourier1} \tilde T^{\mu\nu}_{\eff}(k)&=  i\sum_{n=1}^2
\e^{\pm i(kb)/2}\biggl(\frac{\tilde P_n^\mu \tilde P_n^\nu}{\tilde
E_n} \frac{1}{\omega-{\bf k} \tilde{\bf v}_n} - \frac{P_n^\mu
P_n^\nu}{E_n} \frac{1}{\omega-{\bf k}{\bf v}_n} \biggr).
\end{align}

To leading order in our approximation the scattering process, we
are dealing with, is elastic with $\tilde{E}_n=E_n=E$.
Write for the incoming particles $P_n^\mu = E u_n^\mu
= E (1,0,0,\pm1)$ and for the outgoing ones $\tilde P_n^\mu=E
\tilde u_n^\mu = E (u_n^\mu + \!\un \dot z_n^\mu)=E(u_n^\mu \mp
\alpha\, \hat b^{\kn\mu}) \equiv P_n^\mu +\! \un P_n^\mu$,
substitute into $\tilde T_{\rm eff}^{\mu\nu}$, and expand in
powers of $\alpha$ to have $|k \cdot\! \un P_n^\mu| \ll |k  \cdot \! P_n^\mu|$, to
obtain
the tensor which is {\it identical} to the local part $T_{\mu\nu}$ we use, at \emph{any}  frequency.

 Therefore, it is natural to impose the angular cutoff $\vartheta>\alpha$
for all frequencies of the emitted graviton.

\medskip

\textbf{Regime VI.} For $a\gg 1$ the two radiation amplitudes read:
\begin{align}
& \tau_{+}(k)\approx -\frac{64 {\kern 1pt}G  e^2}{\sqrt{2} a^2}\,
 \cos 2 {\varphi}  \cos \frac{a\cos \varphi}{2}\,, \qquad
 \tau_{\times}(k) \approx  -\frac{64 i{\kern 1pt} G e^2}{\sqrt{2} a^2}\,
 \sin 2
 \varphi\, \sin
 \frac{a\cos \varphi}{2}\,,\nn
\end{align}
from which one can obtain an estimate for the frequency distribution of the emitted radiation in regime VI by integrating over $\varphi$ and over $\vartheta \in (\alpha, \pi-\alpha)$, namely
\begin{align}\label{dw}
\frac{dE^{V\!I}_{\rm rad}}{d\omega }  \sim \frac{\alpha E }{
b} \,\frac{1}{\omega^{2} } \,, \qquad \omega>
1/\alpha b\,,
\end{align}
as well as an estimate for the emitted energy and the corresponding efficiency in regime VI, by integrating also over $\omega\in (1/\alpha b, \infty)$,
\begin{align}
\label{Erad} E^{V\!I}_{\rm rad}\sim \alpha^2 E  \qquad {\rm and}
 \qquad \epsilon_{V\!I} \sim \alpha^2.
\end{align}

\medskip

\textbf{Regime IV.} Inside the regime IVa the amplitude is damped as in
regime VI. However, near the left border of regime IVb (with
$1/b\lesssim \omega \ll 1/\alpha b$) one may expand the amplitudes
in powers of $a$ and obtain (\ref{scalampl2dd}).
Upon integration over regime IVb, i.e. for $\alpha \lesssim \vartheta \lesssim \vartheta_{\rm max}=\arcsin(1/\omega b)$, one obtains
 \begin{align}
 %\label{fr_di2k}
\left( \frac{dE^{I\!V}_{\rm rad}}{d\omega} \right)_{\!1/b\lesssim
\omega\ll 1/\alpha b}\simeq
 \frac{\alpha^3  E\kn b}{\pi} \,\ln \frac{ \tg(\vartheta_{\max} /2)}{\tg(\alpha/2)}
\simeq \frac{\alpha^3  E \kn b}{\pi}\,\ln \frac{2 \alpha^{-1}
}{\omega b + \sqrt{\omega^2 b^{\kn 2} -1}}\,. \nn
\end{align}
Thus, for $1/b\lesssim \omega \ll 1/\alpha b$ one may approximate
$dE_{\rm rad}/d\omega$ by
\begin{equation} \left( \frac{dE^{I\!V}_{\rm rad}}{d\omega}
\right)_{\!1/b\lesssim \omega\lesssim 1/\alpha b} \sim  \alpha^3
E\kn b\, \ln \frac{2}{\alpha \omega b} \,.
\end{equation}
It should be pointed out here that the integral of $dE_{\rm
rad}/d\omega$ over $\omega$ receives most of its contribution from
frequencies in the neighborhood of $1/\alpha b$ in both regimes IV
and VI. Thus, one can say that the characteristic frequency of the
emitted radiation is around ${\mathcal O}(1/r_S)$.

\medskip

\textbf{Total emitted radiation.} Upon integration (\ref{scalampl2dd}) leads to the angular distribution
\begin{align} \label{cher6}
 \frac{dE_{\rm rad}}{d \vartheta}=\frac{\eta \alpha^3 E}{8\pi^2}
 \frac{1}{\sin^2\!\vartheta}\,.
\end{align}
Integration over $ \vartheta\in (\alpha,\pi-\alpha)$ gives
\begin{align} \label{cher7}
 E_{\rm rad} =\frac{\eta {\kern 1pt}\alpha^2 E}{4\pi^2}\,,\qquad\qquad
\epsilon =\frac{ E_{\rm rad}}{2E} \simeq 1.14\,\alpha^2\,.
\end{align}

\section{Conclusions -- Discussion}

Using the same approach as in \cite{GST}{}, based on standard GR,
with the leading non-linear gravity effects taken into account, we
studied collisions of massless particles and computed the
gravitational energy of arbitrary frequency, which is emitted
outside the cone of angle $\alpha =2 r_S/b \ll 1$ in the forward
and backward directions. The value $\epsilon \simeq 1.14\,
\alpha^2$ was obtained for the radiation efficiency, with
characteristic frequency $\omega \sim 1/r_S$.  In fact, this value
represents a lower bound of the efficiency, since it does not
include the energy emitted inside that cone. The frequency
distribution of radiation in the characteristic angle-frequency
regimes is shown in Fig.\,\ref{table}. Unfortunately, we cannot
yet confirm the presence of any other characteristic frequency\cite{GV}{},
such as e.g. $1/\alpha^3 b$, or characteristic emission angle
smaller than $\alpha$. We hope to return to these issues
with a better understanding of regime V  in the near future.

\bigskip
{\bf Acknowledgments.}  We would like to acknowledge enlightening discussions with D.\,Gal'tsov,
G.\,Veneziano and D.\,Colferai. This work was supported in part by the EU program ``Thales" (MIS 375734) as well as by the RFBR grant 14-02-01092. PS acknowledges the non-commercial ``Dynasty''
Foundation (Russian Federation) for financial support. The work of TNT is implemented under the ``ARISTEIA II" Action of the Operational Program ``Education and Lifelong Learning" and is co-funded by the European Social Fund (ESF) and Greek National Resources and was also partially supported by the European Union Seventh Framework Program (FP7-REGPOT-2012-2013-1) under grant agreement No. 316165.

\end{document}